\documentclass[twocolumn,aps,showpacs,pra]{revtex4}
\usepackage{epsfig}
\usepackage[english]{babel}
\usepackage{latexsym}
\usepackage{graphics}
\usepackage{subfigure}
\usepackage{epsfig}
\usepackage{graphicx}
\usepackage{dcolumn}
\usepackage{amsmath}
\usepackage{hyperref}

\begin{document}
\title{Classical effect for enhanced high harmonic yield in  ultrashort laser pulses \\ with a moderate laser intensity}
\author{Y. Z. Shi$^{1}$,  F. L. Dong$^{2}$, Y. P. Li$^{1}$, S. Wang$^{1}$, and Y. J. Chen$^{1*}$}

\date{\today}

\begin{abstract}
We study the influence of the pulse duration on high harmonic generation (HHG) with exploring a wide laser-parameter region theoretically.
Previous studies have showed that for high laser intensities near to the saturation ionization intensity,
the HHG inversion efficiency is higher for shorter pulses
since the ground-state depletion is weaker in the latter.
Surprisingly, our simulations show this high efficiency also appears even for a moderate laser intensity at which the ionization  is not strong.
A classical effect relating to shorter travel distances of the rescattering electron in shorter pulses, is found to
contribute importantly to this high efficiency.
The effect can be  amplified significantly as a two-color laser field is used,
 suggesting an effective approach for increasing the HHG yield.

\end{abstract}
\affiliation{1.College of Physics and Information Technology,
Shaan'xi Normal University, Xi'an, China\\ 2.College of Physics and Information Engineering,
Hebei Normal University, Shijiazhuang, China} 
\maketitle

\section{Introduction}
Because of the promising application as the attosecond light source\cite{Philippe,Corkum2,Corkum3},
high harmonic generation (HHG) has attracted great  interests in  recent years \cite{Kapteyn4,Mairesse,Itatani}.
According to the well-known three-step model \cite{Corkum},
the maximal energy of the  harmonic (the cutoff energy)  in the HHG  is $I_p+3.17U_p$.  Here, $U_p=E_0^2/(4\omega_0^2)$ is  the ponderomotive
energy with $E_0$ and $\omega_0$ being the laser
amplitude and  frequency.  $I_p$ is the ionization potential of the ground state.
To obtain shorter attosecond pulses, higher cutoff energy
is expected. To increase the HHG cutoff, one can increase the laser intensity or the wavelength.
However, there are limitations for both manipulations.
First, high laser intensities can induce the important depletion of the ground state which decreases the HHG yield \cite{Christov,Ditmire}.
Secondly, due to the diffusion effect, the HHG yield also
decreases very fast as the laser wavelength increases \cite{Tate,Shiner,Anh-Thu}.
This decrease of the HHG yield also results in the decrease of the HHG conversion efficiency \cite{Gkortsas}, limiting the
wider application of the HHG.

To overcome these difficulties, great efforts have been devoted \cite{Nam,Yakovlev,Lan}.
It has been found that the target atom can survive higher laser intensities in  ultrashort laser pulses,
resulting in much higher HHG cutoff energy and yield  at high laser intensities near to
the saturation ionization intensity \cite{Kapteyn3,Brabec,Nisoli,Piraux,Krausz2}.
In addition, the use of two-color laser fields has been shown to be a very effective approach
for increasing the HHG cutoff and producing brighter and shorter attosecond pulses \cite{Mauritsson,Dudovich,Zhinan,Mashiko,Misha}.

The motivation of the paper is  to further explore the procedure which could increase the HHG yield in a wide parameter region theoretically.
We revisit the influence of the pulse duration on the HHG with varying the laser intensity and wavelength and working at both one-color and two-color laser fields.
Unexpectedly, our simulations show that  even for a moderate laser intensity with the low ionization probability,
the HHG efficiency still increases remarkably in an ultrashort laser pulse.
Our analyses reveal  a classical effect, which affects importantly on this phenomenon:
during the fast falling part of the short pulse, the rescattering electron
is capable of obtaining the same energy with traveling a shorter distance and therefore enjoys
a more efficient recollision for the HHG than it does in the long pulse.
This classical effect becomes more remarkable as a two-color ultrashort pulse is used with
increasing the HHG conversion efficiency significantly at diverse laser wavelengthes.
Our findings have important implications on the dynamics of the electron in strong ultrashort laser pulses.

The paper is organized as follows.  In Sec. II, we introduce our theoretical methods.
In Sec. III, we show our two-dimensional (2D) numerical results for enhanced HHG efficiency in short laser pulses as the laser intensity is relatively low.
The classical effect responsible for enhancing the HHG yield is discussed in Sec. IV.
 Extended discussions for three-dimensional (3D) cases, for higher laser intensities and different forms of the laser envelope are presented in Sec. V.
Sec. VI is our conclusion.
\begin{figure}[t]
\begin{center}
\rotatebox{0}{\resizebox *{8.5cm}{8cm} {\includegraphics {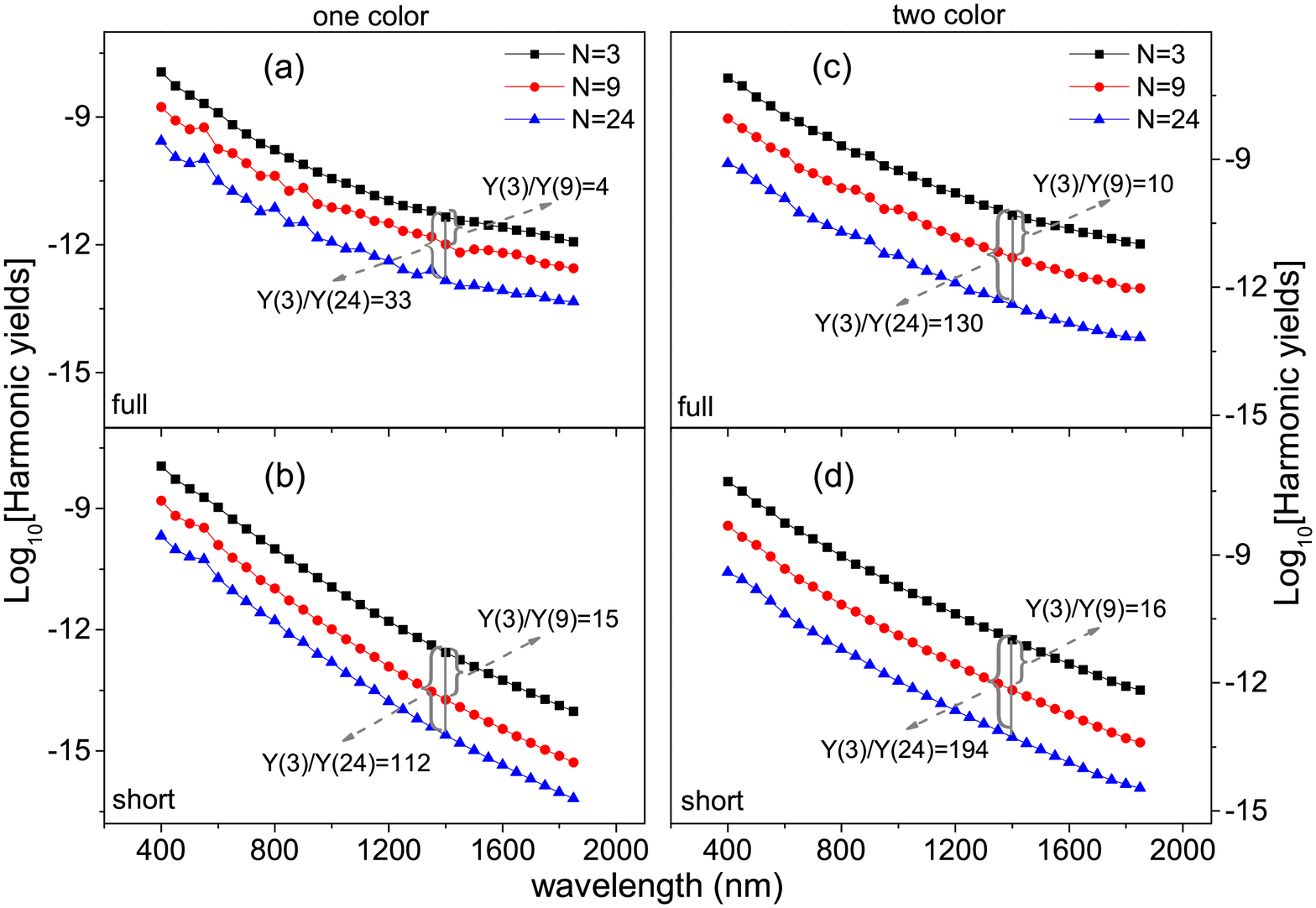}}}
\end{center}
\caption{(Color online)
Wavelength dependence of the  HHG yield  $Y(N,\lambda)$
for $N$-cycle  one-color (a, b) and two-color (c, d) laser pulses.
Results are obtained through full TDSE simulations (a, c) and the  short-trajectory simulations (b, d).
The ratio  of $Y(N_1)/Y(N_2)$   at $\lambda=1400$ nm is also shown.}
\label{fig.1}
\end{figure}

\section{Theoretical methods}
To explore a wide range of  laser wavelength, we first use
a  2D H$_{2}^{+}$ model, which has  been widely used in theoretical studies \cite{Lein}, to simulate the HHG.

The Hamiltonian of the model molecule studied here is
H$(t)=\mathbf{p}^2/2+V(\mathbf{r})+\mathbf{r}\cdot \mathbf{E}(t)$
(in atomic units of $\hbar=e=m_e=1$).   The
potential used here has the  form of
$V(\mathbf{r})=-{Z}/{\sqrt{\xi+r_{1}^2}}-
{Z}/{\sqrt{\xi+r_{2}^2}}$ with
$r_{1,2}^2=({x\pm R/2\cos\theta})^{2}+(y\pm R/2\sin\theta)^2$. 
 $R=2$ a.u. is the internuclear distance,  $\xi=0.5$ is the smoothing parameter which is used to
avoid the Coulomb singularity. $Z$ is the effective charge which is modulated in such a manner that  the ground state  of the model molecule
has the ionization potential of $I_p=1.1$ a.u.. The latter is somewhat higher than that of the He atom.
 $\theta$ denotes the angle
between the molecular axis and the laser polarization.
Here, we have assumed that the laser polarization is along the $x$ axis and the molecular axis is located at the $xoy$ plane.
We consider the perpendicular orientation with  $\theta=90^0$ for which  
 the molecule behaves similarly to an atom \cite{Chen}.
We  work with a space grid size of $L_x\times L_y=1638.4\times 102.4$ a.u. for the $x$  and the $y$ axes.
The electric field  used here has the form of   ${E}(t)=f(t){E_0}(\sin{\omega_{0}}t+\varepsilon\sin{2\omega_{0}}t)$
with $\varepsilon=0$ for one-color cases and $\varepsilon=0.5$  for two-color cases.
$f(t)$ is the envelope function.
To check the influence of the pulse duration on the HHG and simplify our discussions,
we use  a $3n$-cycle 
laser pulse which is switched on and off linearly over $n$ optical
cycles with  $n=1,3,8$. The whole pulse duration is $NT$ with $N=3n$ and $T=2\pi/\omega_0$.

We solve the time-dependent Schr\"{o}dinger
equation (TDSE)  using the spectral method.
In each time step, a mask function $\cos^{1/8}$ is used in the boundary to absorb the continuum wave packet. The coordinate position $x_0$ ($y_0$) from where
the mask function becomes to work is $\pm L_{x}/8$ ($\pm L_{y}/8$). Alternatively, we can set  $x_0=\pm E_0/w_0^2$
with $y_0=\pm L_y/8$ unchanged \cite{Yu}. For our present cases, this treatment removes the contributions of the long trajectory and
multiple returns to the HHG as the contribution of the short trajectory,  which dominates experimental HHG \cite{Agostini},
is not influenced basically. $E_0/w_0^2$ is the quiver amplitude of the classical electron in the laser field.
Below, for differentiation from the full TDSE simulation with  $x_0=\pm L_{x}/8$, we denote the simulation with setting  $x_0=\pm E_0/w_0^2$ as the short-trajectory simulation.

Once the  HHG power spectrum $S(\omega)$ for the harmonic $\omega$ is obtained from the TDSE dipole acceleration,  we calculate the average HHG yield for a $N$-cycle laser pulse
using $Y(N,\lambda)\propto\frac{1}{N(\Omega_{2}-\Omega_{1})}\int ^{\Omega_{2}}_{\Omega_{1}} S(\omega) d\omega$.
Here, $\Omega_1=I_p$ and 
$\Omega_2$ is the cutoff energy of the  spectrum.
$\lambda$ is the laser
wavelength. For the  pulse shape used here, $Y(N,\lambda)$ can be used to compare the HHG conversion efficiency at different pulse durations directly \cite{Gkortsas}.

In the following,  our discussions will be performed for a moderate laser intensity of $I=5\times10^{14}$W/cm$^{2}$
at which the ionization yield of the  model molecule is low.

\begin{figure}[t]
\begin{center}
\rotatebox{0}{\resizebox *{8.5cm}{8cm} {\includegraphics {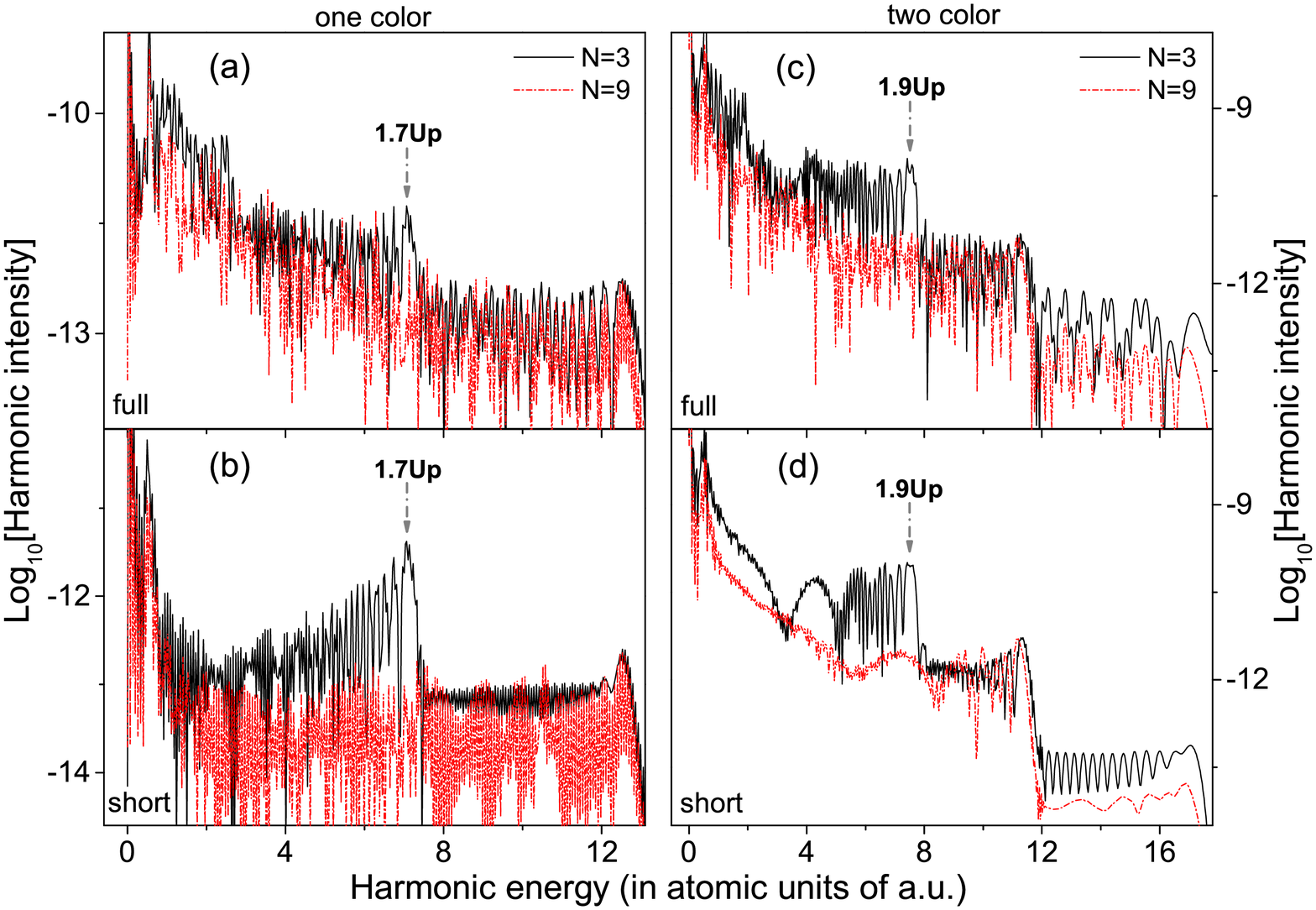}}}
\end{center}
\caption{(Color online) HHG spectra
for $N$-cycle  one-color (a, b) and two-color (c, d) laser pulses at $\lambda=1400$ nm. Results are obtained through full TDSE simulations (a, c)
and the  short-trajectory simulations (b, d), and divided by the  cycle number $N$.}
\label{fig.2}
\end{figure}

\section{Enhanced HHG yield}
In Fig. 1(a), we plot the  wavelength dependence of the average HHG yield in a one-color field for
different pulse durations. The contrast of the curves is remarkable. One can observe that the average HHG yield
is the highest for the 3-cycle pulse at different wavelengthes. This yield decreases as the cycle number $N$ increases. At long wavelengthes such as
$\lambda=1400$ nm, the  yield of the 3-cycle pulse is several times higher than the 9-cycle result,
and one order of magnitude higher than that of the 24-cycle pulse. Note, in this case, the whole HHG
yield of $NY(N,\lambda)$ for $N=3$ is also  several times higher than that for $N=24$.
The contrast of the HHG yields at
different pulse durations becomes more remarkable
as the short-trajectory simulations are performed.
 As shown in Fig. 1(b), as the contributions of the long trajectory and multiple returns are excluded,
 the short-trajectory HHG yield for $N=3$ is one order of magnitude higher than $N=9$, and
two orders of magnitude higher than $N=24$. For a two-color laser field, however, this remarkable difference occurs even for the full TDSE simulations,
as shown in Fig. 1(c). Here,
the ratio of $Y(3,\lambda)$ vs $Y(24,\lambda)$ with $\lambda=1400$ nm arrives at $130$, implying
a significant increase of the HHG efficiency in a short two-color laser pulse. This significant increase of the efficiency
is further amplified as the short-trajectory contributions are considered, as  shown in Fig. 1(d).
Since the short-trajectory simulations are closely associated with  the  classical motion of the electron,
these results in Figs. 1(b) and 1(d)  imply that the potential mechanism
which   increases the HHG yields at short pulses, is  related to
the classical aspect of the electron. Next, we explore the mechanism in detail.

\begin{figure}[t]
\begin{center}
\rotatebox{0}{\resizebox *{8.5cm}{8cm} {\includegraphics {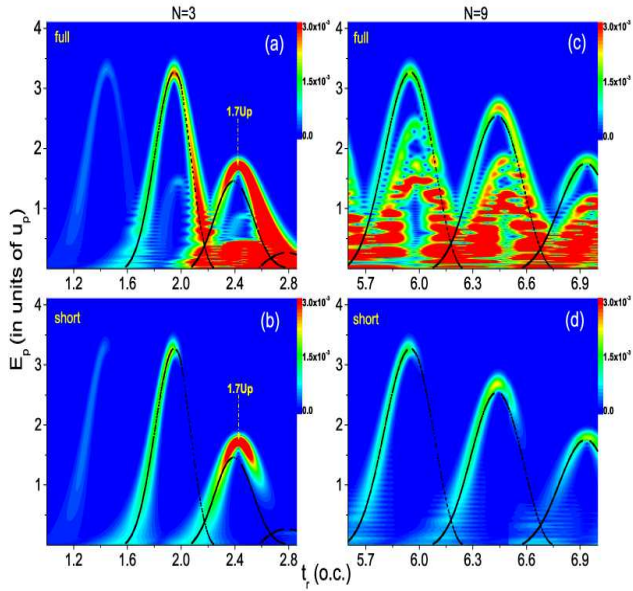}}}
\end{center}
\caption{(Color online) Rescattering time and energy distributions
(the color coding)
for 3-cycle (a, b) and 9-cycle (c, d)  one-color laser pulses at $\lambda=1400$ nm.
Results are obtained through full TDSE simulations (a, c)
and the  short-trajectory simulations (b, d).
In each panel,
the black-square curve
shows
the electron trajectory for the first return, obtained from the quantum orbit model.
}\label{fig.3}
\end{figure}

Figure 2 plots the HHG spectra of 3-cycle vs 9-cycle pulses  for the typical wavelength of $\lambda=1400$ nm. 
For comparison, these spectra are divided by the cycle number $N$.
For the one-color case in the left column of Fig. 2, one can observe from Fig. 2(a) of full  simulations:
i) the spectrum with $N=3$ (the solid-black curve) is
higher than that of $N=9$ (the dashed-red curve), especially for the low-energy part. ii) In both cases,
the HHG yield decreases as the harmonic energy increases. iii)
The spectrum of $N=3$ shows three plateaus with the cutoff positions of
$\omega=2.5$ a.u., $7.1$ a.u. and $12.5$ a.u. respectively. Around the  second cutoff $\omega=7.1$ a.u.
(corresponding to the  electron kinetic energy of $E_p=\omega-I_p=1.7U_p$),  a robust peak
can be observed. As
the short-trajectory simulation is executed, the robust peak of $N=3$ survives our treatments and the spectrum of $N=9$ becomes flat,
resulting in a remarkable contrast of these two spectra with different $N$, as
shown in Fig. 2(b).

For the two-color case, the contrast of the two spectra at different pulse durations is more remarkable, even for the full TDSE results, as shown in the right column of Fig. 2.
In this case, the spectrum of $N=3$ in Fig. 2(c) of full simulations shows four plateaus with the
cutoff positions of $\omega=1.9$ a.u., $7.6$ a.u., $11.3$ a.u. and $17.1$ a.u., respectively.
Around the second plateau which also has a robust peak at $\omega=7.6$ a.u. (corresponding to $E_p=1.9U_p$), the spectrum of $N=3$
is one order of magnitude higher than that of $N=9$. The latter, by comparison, shows two striking plateaus corresponding to the third and the fourth plateaus of $N=3$.
As we perform the short-trajectory simulations,
the intensity of the second plateau does not decrease basically,
as shown in Fig. 2(d). The origin of the plateaus can be well understood through the wavelet analysis of  the TDSE dipole acceleration
combined with the quantum orbit (QO) theory \cite{Lewenstein,Salieres},
as to be discussed below.
These comparisons in Fig. 2 explain the remarkable difference for the  HHG yields at different pulse durations discussed in Fig. 1. From the comparisons,
one can also conclude that the high HHG efficiency of short pulses is closely related to the robust HHG peaks of $1.7U_p$ and $1.9U_p$ indicated in Fig. 2, which are
less influenced by our different absorbing procedures in TDSE simulations. This conclusion can  be further checked through the  time-frequency analysis.

\section{Classical effects}
As a case, in Fig. 3, we show wavelet-analysis  \cite{Tong} results (the color coding) for the corresponding spectra  in Figs. 2(a) and 2(b).
For comparison, here, we also show the electron trajectory  of the first return with the excursion time of the electron shorter
than a laser cycle (the black-square curve),
predicted from the QO model.
For $N=3$, from  Fig. 3(a)  of the full TDSE results,  one can observe that i) the distributions  match the theory predictions basically;
ii) the distributions imply three HHG cutoffs. The first one with $E_p=0.5U_p$ around  the return time $t_r=2.4T$
has the largest amplitude. The second one with $E_p=1.7U_p$ also appears around  $t_r=2.4T$ and has a comparable amplitude with the first one, as indicated by the dashed arrow.
The third one with $E_p=3.2U_p$ near to $t_r=2T$ has the smallest amplitude.
The first one can be identified as arising from multiple returns, and the second and the third ones come from the first return.
For short-trajectory simulations in Fig. 3(b),   the first one  disappears,
as the second and
the third ones keep their amplitudes with the prevailing role of the second one.
Note, the contributions of the long trajectory around  the minus-chirp part of the black-square curve also disappear basically  due to the absorbing procedure used here.
For $N=9$, the situation is different. The full TDSE results in Fig. 3(c) show large amplitudes around the electron trajectories of
multiple returns (these trajectories are not shown here), in agreement with the results in Ref. \cite{Tate}.
In Fig. 3(d) of short-orbital simulations, the contributions of multiple returns disappear. The survived distributions, however, show  smaller
amplitudes  than that around $E_p=1.7U_p$  in Fig. 3(b). To understand the large amplitude located at  $E_p=1.7U_p$ in Fig. 3(b), in Fig. 4, we further compare
the maximal displacement $x_m$ \cite{Yu},
which the electron can travel as it ionizes at the time $t_i$ and returns at $t_r$ for $N=3$ vs $N=9$.
\begin{figure}[t]
\begin{center}
\rotatebox{0}{\resizebox *{8.5cm}{8cm} {\includegraphics{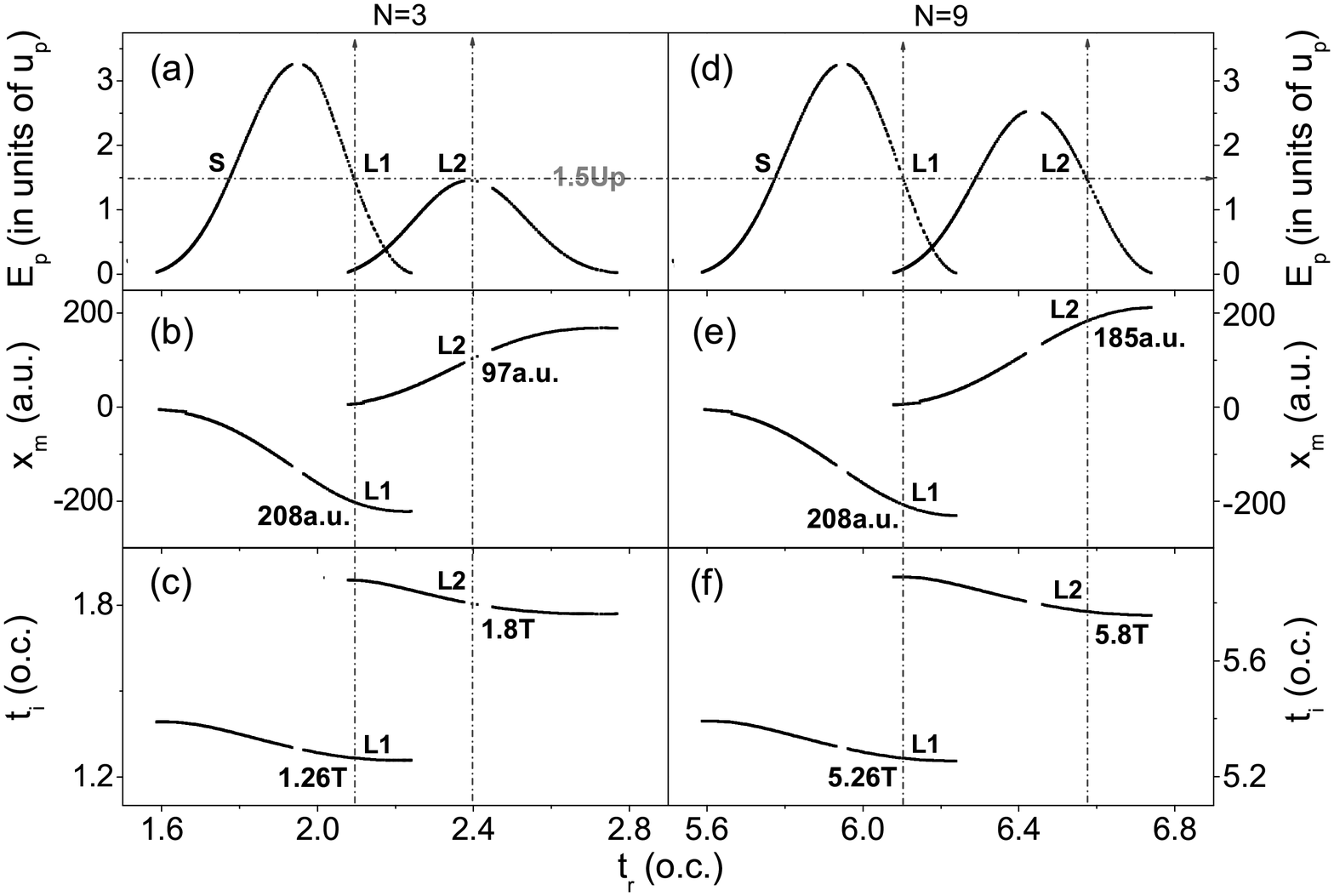}}}
\end{center}
\caption{
The  kinetic energy $E_p$, the maximal displacement $x_m$ and the ionization time $t_i$ of the rescattering electron as functions of the return time $t_r$
for 3-cycle (the left column) and 9-cycle (the right column)  one-color laser pulses at $\lambda=1400$ nm,  obtained using the quantum orbit model.
Only the first return with the excursion time $\tau=t_r-t_i<T$ is shown.
The horizontal arrow indicates the  orbits with  $E_p=1.5U_p$ for short (S) and long (L1 and L2) trajectories. The vertical arrows indicate
the corresponding  $x_m$ and  $t_i$ of theses two long trajectories. The relevant values of $x_m$ and $t_i$ are as shown.}
\label{fig.4}
\end{figure}

\begin{figure}[t]
\begin{center}
\rotatebox{0}{\resizebox *{8.5cm}{8cm} {\includegraphics{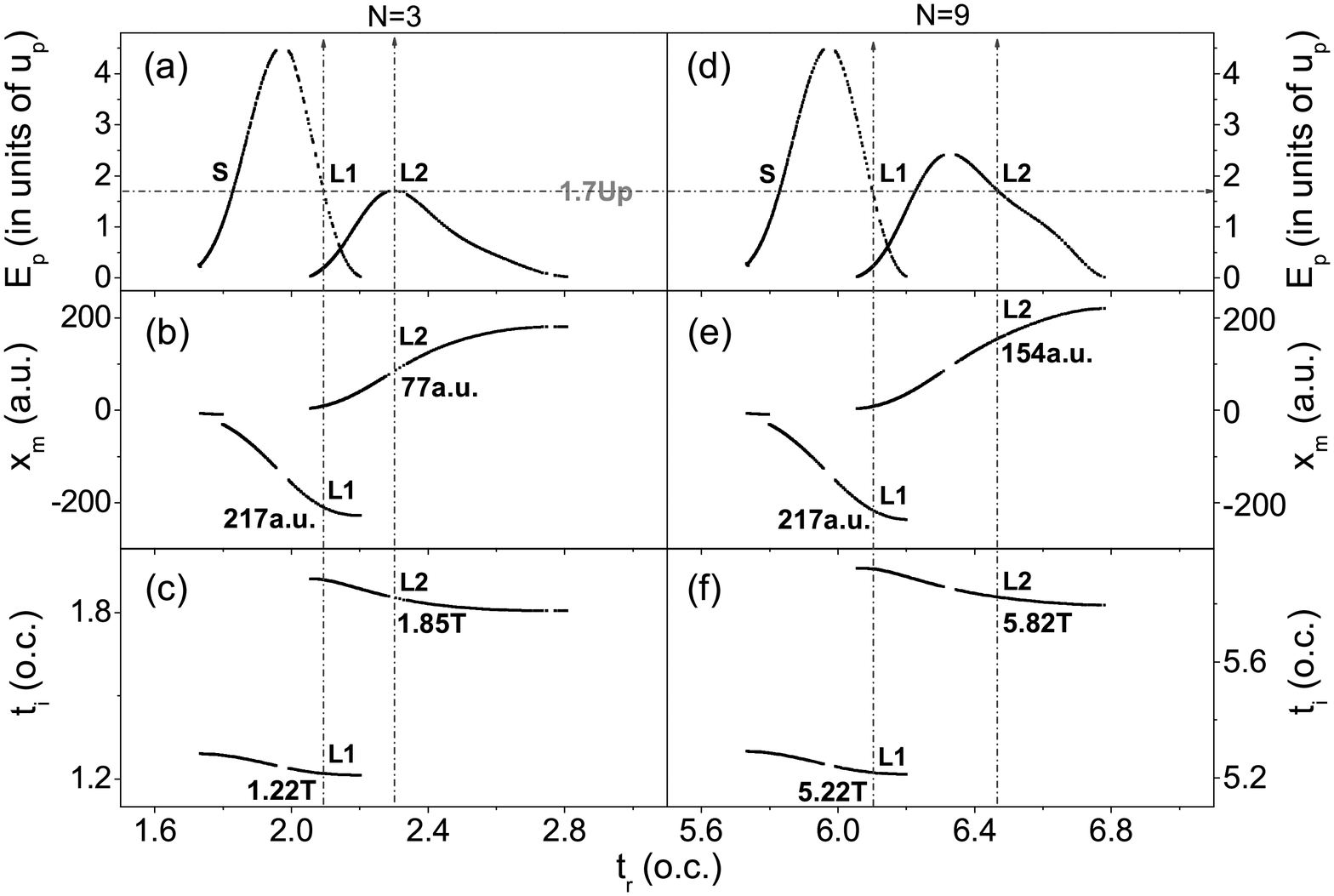}}}
\end{center}
\caption{ Same as Fig. 4, but for two-color laser pulses.}
\label{fig.5}
\end{figure}

To obtain the maximal displacement  $x_m$, we first calculate the complex ionization time $t^q_i$ and the return time $t^q_r$ (the real parts of the complex times are considered
the physical ionization time $t_i$ and the return time $t_r$) by the QO model. Then we evaluate the maximal displacement using  \cite{Becker2}
$x_m\equiv x_m(t)=(E_0/\omega_0^2)Re[\sin\omega_0t-\sin\omega_0t_i^q+p_{st}(\omega_0^2/E_0)(t-t_i^q)]$ with $v(t)=Re[p_{st}+E_0/\omega_0\cos(\omega_0t)]=0$ and $Re(t_i^q)<t<Re(t_r^q)$.
$v(t)$ is the electron velocity,
and $p_{st}=(E_0/\omega_0)[\sin\omega_0t_i^q-\sin\omega_0t_r^q]/[\omega_0(t_r^q-t_i^q)]$ is the saddle-point momentum.
We mention the maximal displacement  obtained here
agrees well with that obtained using the classical procedure introduced in Ref. \cite{Yu}.

Our comparisons are performed for two typical long trajectories (denoted using $L1$ and $L2$ in Fig. 4) with the same return energy $E_p=1.5U_p$,
near to $E_p=1.7U_p$ of the large amplitude in Fig. 3(b).  Both these two trajectories have
the ionization times $t_i$ located in the flat-top part of the pulse and near to the peak of the field, as shown in Figs. 4(c) and 4(f), and  are expected to
contribute importantly to the HHG. The return times of the two trajectories are different.
As the $L1$ trajectory returns near to the flat-top part of the pulse, the $L2$ trajectory returns in the falling part of the pulse (the laser pulse
becomes to fall at 2T for $N=3$ and 6T for $N=9$). For the 3-cycle case in Fig. 4(b),
the displacement $x_m$ for the $L2$ trajectory is 97 a.u.. It is 208 a.u. for  $L1$, two times larger than the $L2$ one. Considering
the spread of the wave packet  is proportional to
the electron displacement,
these above results imply the $L2$ trajectory has a  amplitude several times larger than the $L1$ one, in agreement with the wavelet-analysis result in Fig. 3(b).
For the 9-cycle case in Fig. 4(e), the situation is  different. The $L2$ trajectory has the maximal displacement of $x_m=185$ a.u. which is near to
208 a.u. of the $L1$ one. As a result, the wave packet spreading is comparable for the two trajectories in the 9-cycle case, resulting in  similar amplitudes for them, as
seen in Fig. 3(d). We mention that  short trajectories have the smaller displacements $x_m$ than the corresponding long ones.
However, they usually ionize at a time farther away from the peak of the field and therefore have   smaller amplitudes than the long ones
(this can also be seen from the wavelet-analysis results in Fig. 3). For the reason, we don't discuss them here.

In combination with the distributions in Fig. 3(b) vs Fig. 3(d),
the contrast of the maximal displacements $x_m$ for the $L2$ trajectory in  Fig. 4(b) vs Fig. 4(e) suggests that
the higher HHG efficiency  for the short pulse of $N=3$, observed in Fig. 1(b), is closely related to
the shorter excursion distance of the electron in the fast falling part of the  short pulse.
This classical effect
which increases the HHG efficiency becomes more remarkable in the two-color case, as shown in Fig. 5. Here, our analyses  are also performed for
two typical long trajectories of $L1$ and $L2$. The ionization times of the two long trajectories both are
located at the flat-top part of the  pulse and are near to the peaks of the two-color field, as shown in Figs. 5(c) and 5(f).
In addition, the electric-field amplitudes at the two ionization times of  $L1$ and $L2$ in the two-color case are nearer to each other
 than in the one-color one. For the short-pulse case of $N=3$,   one can observe from Fig. 5(b) that
the maximal displacement  of the $L2$ trajectory is $x_m=77$ a.u., as the $L1$ trajectory shows a maximal
displacement of $x_m=217$ a.u., almost three times larger than the $L2$ one. For the long-pulse case of $N=9$, they are 154 a.u. and 217 a.u., respectively, as shown in Fig. 5(e).
The shorter excursion distance of the $L2$ trajectory
in Fig. 5(b) of $N=3$, suggests that this trajectory contributes significantly to the HHG in the short-pulse case.
It is corresponding to the cutoff of the second plateau with  high intensity in Figs. 2(c) and 2(d).
We mention that the TDSE cutoff position of the second plateau in Fig. 2(c) is $1.9U_p$ for the 3-cycle case, somewhat higher than the model prediction of $1.7U_p$ in Fig. 5(a). This is also the one-color case of $1.7U_p$ in Fig. 2(a) versus $1.5U_p$ in Fig. 4(a). This difference can partly arise from the nonadiabatic effect in ultrashort pulses.
In comparison with $x_m=97$ a.u. with $E_p=1.5U_p$ in Fig. 4(b), this shorter excursion distance $x_m=77$ a.u. of
the $L2$ trajectory with the higher return energy  $E_p=1.7U_p$  in Fig. 5(b) also suggests the HHG efficiency is higher in the two-color case than in the one-color case, in agreement with
our analyses in Fig. 1.

\section{Extended considerations}

\subsection{Three-dimensional simulations}
To check our results, we have also performed 3D simulations for H$_2^+$ with the soft-core potential of $V(\mathbf{r})=-{Z}/{\sqrt{\xi+r_{1}^2}}-
{Z}/{\sqrt{\xi+r_{2}^2}}$ and  $r_{1,2}^2=({x\pm R/2\cos\theta})^{2}+(y\pm R/2\sin\theta)^2+z^2$.
The Definitions of the parameters  $Z$, $\xi$, $R$ and $\theta$ are the same as in  our 2D cases.
The 3D calculations are very time-memory consuming. Here, we work with a grid size of
$L_x\times L_y\times L_z=819.2\times51.2\times51.2$ a.u. for the $x$, $y$ and $z$ axes,
respectively. Our calculations are performed for  $I=5\times10^{14}$W/cm$^{2}$ and $\lambda=1400$ nm
corresponding to the laser parameters used in Fig. 2. Similar absorbing procedures as in 2D cases
are also used in performing full TDSE simulations and short-trajectory simulations.
The  results are presented in Fig. 6.

\begin{figure}[t]
\begin{center}
\rotatebox{0}{\resizebox *{8.5cm}{8cm} {\includegraphics {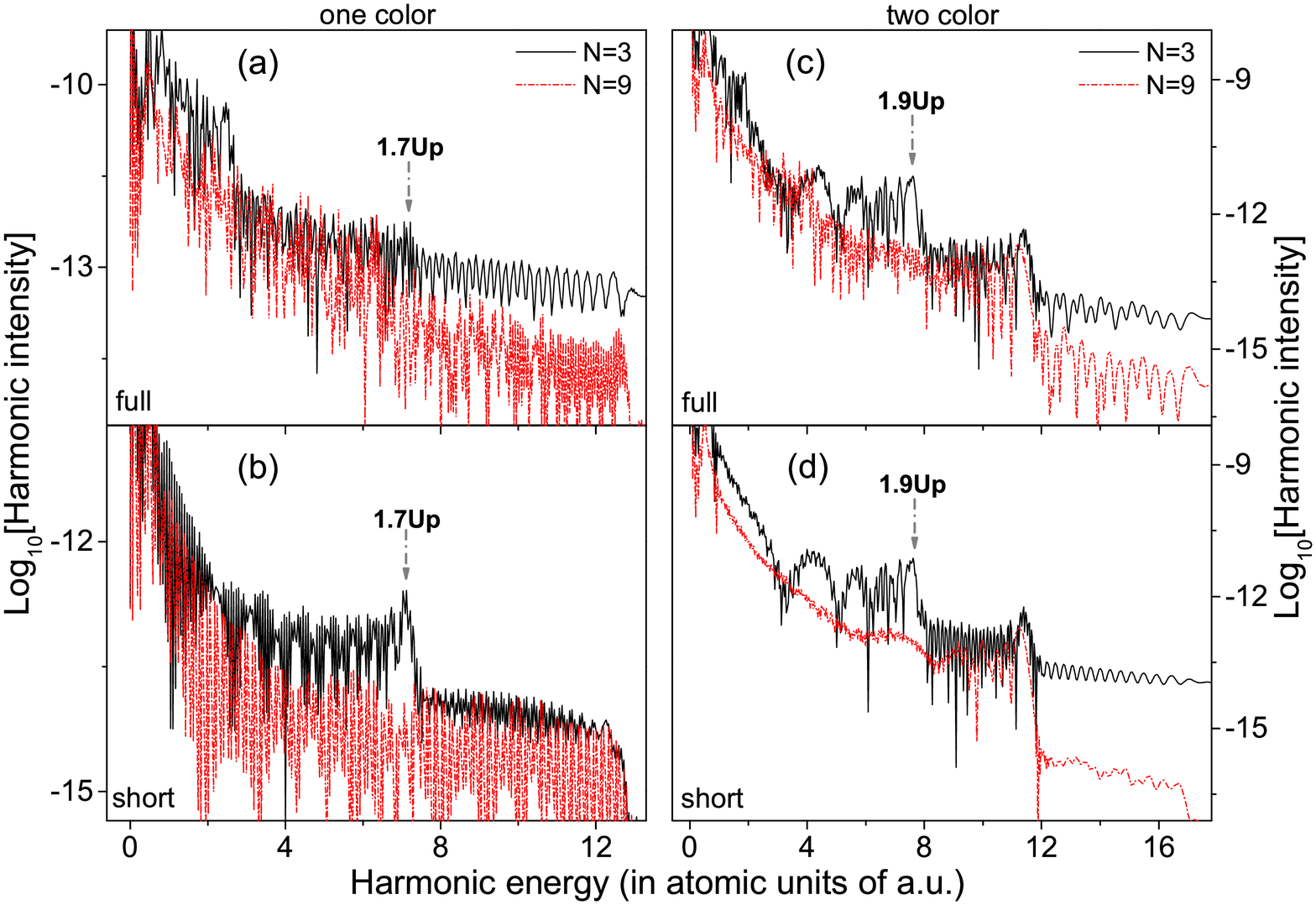}}}
\end{center}
\caption{(Color online) Same as Fig. 2, but obtained with 3D simulations.}
\label{fig.6}
\end{figure}

One can observe from Fig. 6 that the 3D results are
similar to the 2D ones in Fig. 2. First, the spectra of $N=3$ show a robust peak
which appears around $E_p=\omega-I_p=1.7U_p$ for one-color cases in the
left column of Fig. 6 and around $E_p=1.9U_p$ for two-color cases in the
right column of Fig. 6.
Secondly, this robust peak is more remarkable in short-trajectory simulations
than in full simulations and in  two-color cases than in one-color cases.
As discussed in Fig. 2, this robust peak arises from the classical effect, which increases the HHG efficiency
importantly in short pulses. The 3D results in Fig. 6 also show that this HHG efficiency is strikingly  higher in the short pulse than in the lone one.
All of the characteristics are in agreement with our 2D results in Fig. 2.

One of the main differences between 2D and 3D cases is that
the diffusion effect relating the wave packet spreading is stronger in 3D cases than in 2D cases.
This stronger diffusion effect can induce a larger phase difference between the harmonics emitted in different laser cycles, especially
for harmonics  arising from the long trajectory and multiple returns with longer excursion times in the laser field. The interference of the  harmonics emitted
in different laser cycles will decrease the intensity of the HHG spectra and this decrease is more remarkable for long pluses with more cycles.
This can be the reason that the spectrum of $N=9$ in Fig. 6(a) of full TDSE simulations shows a smaller amplitude in comparison with that of $N=3$ in the
high-energy region with $\omega>7$ a.u..  Note that in Fig. 6(b) of short-trajectory simulations, these two spectra of $N=3$ and $N=9$
are comparable for the high-energy region of $\omega>7$ a.u.,
as the contributions of the long trajectory and multiple returns are excluded.

\begin{figure}[t]
\begin{center}
\rotatebox{0}{\resizebox *{8.5cm}{8cm} {\includegraphics {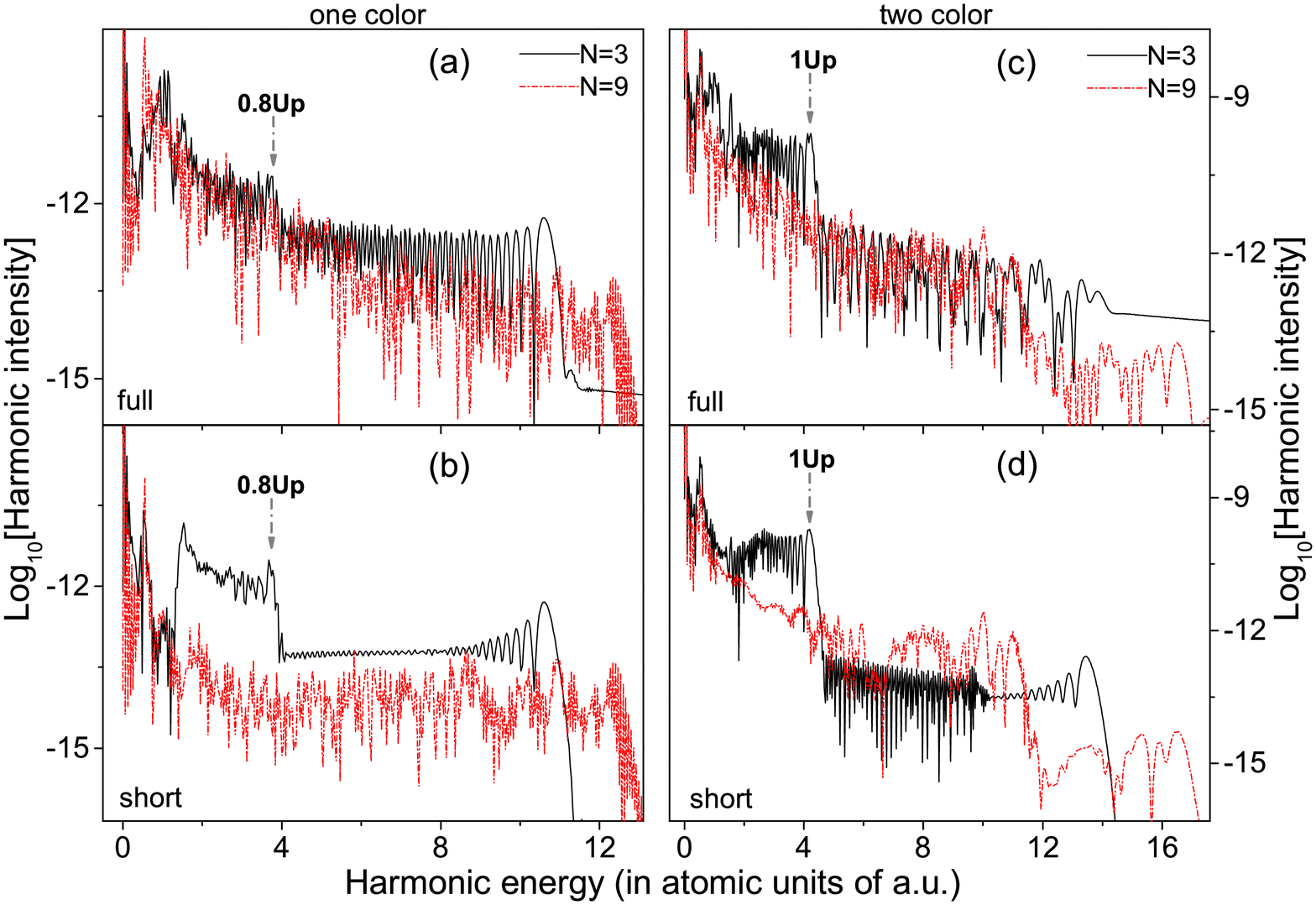}}}
\end{center}
\caption{(Color online)  Same as Fig. 2, but obtained for a sin-square-envelope laser pulse.}
\label{fig.7}
\end{figure}

\subsection{Sin-square-envelope pulses}

To check our results, a sin-square-envelope pulse is also used in our calculations and relevant results are presented in Fig. 7.
The results in Fig. 7  are also similar to those in Fig. 2, with showing a robust peak in
the spectra of $N=3$. Here, the position of the peak  is around $E_p=0.8U_p$ for one-color cases and  $E_p=U_p$ for two-color cases, somewhat lower than those
 in Fig. 2. In addition, the maximal cutoff position of the spectrum of $N=3$ is also somewhat lower than that of $N=9$. However, in comparison with
the results in Fig. 2, the results for $N=3$ presented here show higher inversion efficiency.  For example, in Fig. 7(b),
the spectrum of $N=3$  is higher than that
of $N=9$ in the whole energy region, different from the results in  Fig. 2(b). In addition,
the curve of $N=3$ in Fig. 7(d) also shows a cutoff located at $\omega=13.5$ a.u.. Around the cutoff,
the spectrum of $N=3$ is two orders of magnitude higher than that of $N=9$.
All of the characteristics can be understood by virtue of the  analyses of the quantum orbit and the maximal displacement of the rescattering electron.

\subsection{High laser intensities}

As the laser intensity increases and is near to the saturation intensity, the situation is somewhat different. As shown in Fig. 8(a),
for the case of $I=1.2\times10^{15}$W/cm$^{2}$  and $\lambda=1400$ nm, the HHG spectrum of $N=3$  is higher than
that of $N=9$ in the whole energy region, even the average over the cycle number $N$ is not performed here. This can be understood from the ground-state deletion. For the long pulse of $N=9$,
the ionization is stronger than that of $N=3$, resulting in a significant deletion of the ground state and accordingly a remarkable decrease of the HHG yield.
However, in this case, the classical effect can still be read from the spectrum, which manifests itself as a harmonic peak around $E_p=1.75U_p$,  as indicated by the
dashed-dotted arrow. This peak is more remarkable in short-trajectory simulations, as shown in Fig. 8(b).
In addition, the cutoff position in the spectrum of $N=3$ in Fig. 8(a) seems somewhat larger than that of $N=9$. This phenomenon is clearer
for the case of $I=1\times10^{15}$W/cm$^{2}$ and $\lambda=1600$ nm, as shown in Fig. 8(c). We mention that for the case in  Fig. 8(c),
the ionization is somewhat weaker than that in Fig. 8(a).
Accordingly, the spectra of $N=3$ and $N=9$ are comparable here. In addition,
a harmonic peak around $\omega=1.78U_p$ can also be observed in the spectrum of $N=3$ in Fig. 8(c). This peak is more striking in Fig. 8(d) of short-trajectory simulations.
These results show that
the classical effect, mainly discussed in the paper,
is general in a wide laser-parameter region.

\begin{figure}[t]
\begin{center}
\rotatebox{0}{\resizebox *{8.5cm}{8cm} {\includegraphics {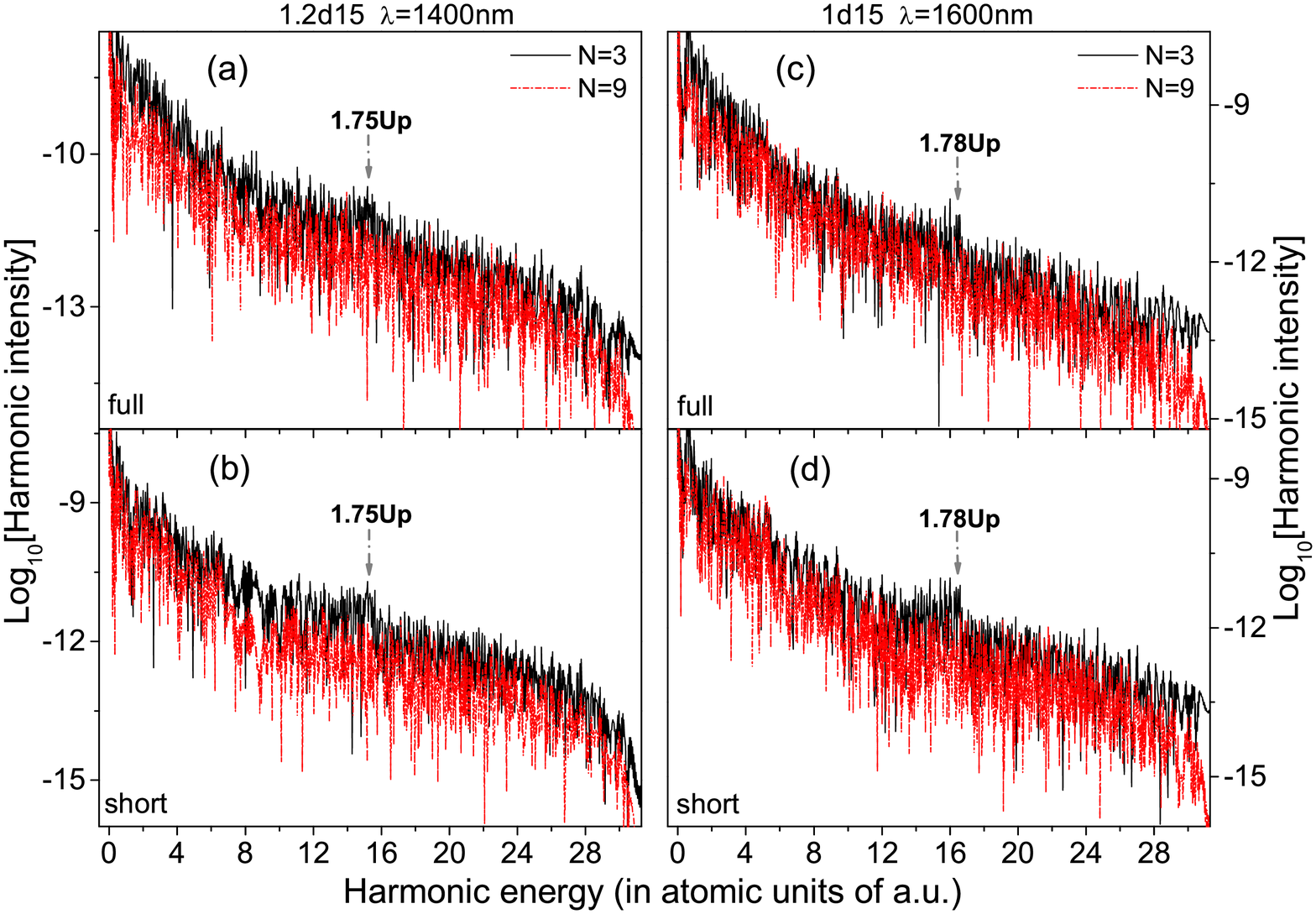}}}
\end{center}
\caption{(Color online) HHG spectra
for $N$-cycle  one-color  laser pulses at  $I=1.2\times10^{15}$W/cm$^{2}$ and $\lambda=1400$ nm (a, b) and
$I=1\times10^{15}$W/cm$^{2}$ and $\lambda=1600$ nm (c, d). Results are obtained through full TDSE simulations (a, c)
and the  short-trajectory simulations (b, d) in 2D cases.}
\label{fig.8}
\end{figure}

\section{Conclusions}
In summary, 
we have shown that the HHG efficiency in an ultrashort laser pulse is influenced significantly by a classical effect. The latter
is closely associated with  the shorter excursion distance of the rescattering electron as it
ionizes near the peak of the short laser  pulse and returns in the fast falling part of the   pulse.
This shorter excursion distance suppresses the spread of the wave packet  and  increases the efficiency of the HHG.
With the shorter excursion distance and
accordingly the shorter excursion time, the emitted harmonics relating to the classical effect can be
easier to survive the macroscopic propagation in the medium and easier to modulate in experiments.
We expect that this effect also has an  important influence on the rescattering induced other strong-field phenomena such as
high-order above threshold ionization and nonsequential double ionization. 

\section*{Acknowledgement}
We thank Professor C. D. Lin for valuable discussions.
This work was supported by the National Natural Science
Foundation of China (Grant No. 11274090) and the Fundamental
Research Funds for the Central Universities (Grant
No. GK201403002).

\end{document}